\documentclass[]{spie}  

\usepackage{amsmath,amsfonts,amssymb}
\usepackage{graphicx}
\usepackage[pdftex,dvipsnames,usenames]{xcolor}
\usepackage{pifont}
\usepackage[colorlinks=true, allcolors=blue]{hyperref}
\usepackage{textpos}

\newcommand{\D}{{\rm{d}}}

\newcommand{\im}{{\rm{Im}}}
\newcommand{\Tr}{{\rm{Tr}}}

\newcommand{\re}{{\rm{Re}}}

\newcommand{\ket}[1]{\left|#1\right\rangle}

\title{Nonclassicality and Bell nonlocality in atmospheric links}

\author[a,b]{A. A. Semenov}
\author[a]{M. Bohmann}
\author[a,c]{D. Vasylyev}
\author[a]{W. Vogel}
\affil[a]{Institut f\"ur Physik, Universit\"at Rostock, Albert-Einstein-Str. 23, D-18059 Rostock, Germany}
\affil[b]{Institute of Physics, NAS of
Ukraine, Prospect Nauky 46, 03028 Kiev, Ukraine}
\affil[c]{Bogolyubov Institute for Theoretical Physics, NAS of Ukraine, Vulytsya
	Metrologichna 14-b, 03680 Kiev, Ukraine}

\authorinfo{Further author information: (Send correspondence to A.A.S.)\\A.A.S..: E-mail: andrii.semenov@uni-rostock.de, Telephone: +49 381-498-6937}

\begin{document}
\maketitle
\begin{textblock*}{7cm}(11cm,-5.2cm)
  \fbox {\footnotesize Proc. of SPIE \textbf{10771}, 107710Z}
\end{textblock*}

\begin{abstract}
	Free-space quantum links have clear practical advantages which are unaccessible with fiber-based optical channels --- establishing satellite-mediated quantum links, communications through hardly accessible regions, and communications with moving objects.
	We consider the effect of the atmospheric turbulence on properties such as quadrature squeezing,  entanglement, Bell nonlocality, and nonclassical statistics of photocounts, which are resources for quantum communications.
	Depending on the characteristics of the given channels, we study the efficiency of different techniques, which enable to preserve these quantum features---post-, pre-selection, and adaptive methods.
	Furthermore, we show that copropagation of nonclassically-correlated modes, which is used in some communication scenarios, has clear advantages in free-space links.
\end{abstract}


\section{INTRODUCTION}
\label{sec:intro}
	
	The study of the transmission of quantum light through atmospheric free-space channels tremendously gained importance over the last decade and developed into the new field of atmospheric quantum optics.
	The interest in atmospheric channels is driven by the goal of establishing a worldwide quantum communication network\cite{hughes17}.
	Such networks use quantum key distribution\cite{gisin02} or quantum state teleportation\cite{bennett93,pirandola15} through free-space links, which can ultimately lead to a quantum internet\cite{kimble08}.
	
	Experimental realizations of quantum-optical free-space links started with the successful implementations of ground-to-ground atmospheric links\cite{ursin07,elser09,heim10,fedrizzi09,capraro12,yin12,ma12,peuntinger14}.
	Soon after, small-scale experiments demonstrated the possibility of satellite mediated applications with quantum light\cite{bourgoin13,nauerth13,wang13,bourgoin15}.
	Only recently, the first experiments with satellites have been reported\cite{vallone15,dequal16,vallone16,carrasco-casado16,takenaka17,liao17,yin17,gunthner17,ren17,yin17b}.
	Yet, the era of atmospheric quantum optics has just started and further rapid developments are to be expected.
	
	In order to optimally exploit the potential of atmospheric quantum channels, it is crucial to gain a profound understanding of such channels and their action on the quantum properties of light.
	Unlike in optical fiber links, atmospheric channels do not exhibit constant losses but fluctuating losses due to the atmospheric turbulence\cite{diament70,perina73,milonni04,berman06,semenov09}.
	The overall action of a fluctuating free-space link on a quantum state can be modeled by a probability distribution of the transmittance (PDT) of the channel\cite{semenov09}.
	This includes the complete description from the sender through the atmosphere to the detection with a finite receiver aperture.
	Various atmospheric PDT models have been developed taking into account different kinds of atmospheric channel conditions\cite{vasylyev12,vasylyev16,vasylyev17,vasylyev18}.
	Note that these models show good agreement with experimental data\cite{usenko12,vasylyev16,vasylyev17}.
	
	Besides the proper understanding of the free-space channels, the study of the quantum properties of light transmitted through such channels is the second important aspect in atmospheric quantum optics.
	In this direction several results have been reported.
	Basic properties of the transmitted light such as the influence of fluctuating losses on the photon statistics\cite{perina73,milonni04} or on squeezing\cite{peuntinger14,vasylyev16} have been studied.
	Different discrete-variable quantum states suffering from atmospheric losses have been analyzed in detail\cite{bohmann15,hosseinidehaj15b,bohmann17d}.
	Furthermore, continuous-variable quantum states are being investigated.
	This includes the study of some entangled Gaussian states\cite{hosseinidehaj15a} and the full characterization of entangled two-mode Gaussian state in turbulent channels\cite{bohmann16}.
	The treatment of Gaussian states has further been extended to single and multi-mode nonclassicality and entanglement conditions\cite{bohmann17a}.
	In addition, the influences of free-space links on the violation of Bell inequalities has been examined\cite{semenov10,gumberidze16}.
	Besides, fluctuating loss channels can be simulated in laboratory experiments\cite{pereira13,bohmann17c} which allows to test specific channels before performing a full-scale experiment through the atmosphere.
	
	In this contribution, we provide a rigorous analysis of Bell nonlocality in atmospheric channels, for which we distinguish the cases of counter- and co-propagating radiation modes.
	We demonstrate that pre-selection techniques, i.e. the selection of events with high transmittances by testing the channels with intense pulses, allow for successful violations of Bell inequalities even under the conditions of intense stray light.
	Our analysis shows, that  correlations of the transmittance efficiencies automatically improve the value of the Bell parameter in the case of co-propagation.
	Furthermore, Gaussian entanglement and quadrature squeezing are other phenomena for which we analyze the influence of atmospheric turbulence.
	For these fundamental quantum phenomena, we identify strong influences of the initial coherent displacements, which cannot be observed in Gaussian channels, including constant attenuations.
	Additionally, we derive the counter-intuitive result that strong squeezing is not useful and even harmful for the transmission of Gaussian entanglement through free-space links.
	We also propose a scheme of adaptive channel correlations, which enables to preserve Gaussian entanglement for the price of additional losses.
	We are convinced that our results will find their application in developing and improving quantum communication systems in atmospheric channels.
	
	The paper is structured as follows. In Sec. \ref{Sec:IOR} we present the most general form of input-output relations for quantum light in atmospheric channels.
	These input-output relations are applied for different nonclassical effects in subsequent sections.
	Sub-Poissonian statistics of photocounts at the receiver site is considered in Sec. \ref{Sec:PCS}.
	In Sec. \ref{Sec:Bell} we have analyzed violation of Bell inequalities for light passing through the atmosphere.
	Transferring of quadrature squeezing is considered in Sec. \ref{Sec:Sqeez}.
	Analysis of distribution of the Gaussian entanglement is presented in Sec. \ref{Sec:Entanglement}.
	In Sec. \ref{Sec:Concl} we give summary and concluding remarks.

\section{INPUT-OUTPUT RELATIONS FOR ATMOSPHERIC CHANNELS}	
\label{Sec:IOR}

	From the point of quantum optics, atmospheric channels can be considered as linear-loss channels.
	This means that they can be described by standard input-output relations between the annihilation operators of the input and output modes, $\hat{a}_\mathrm{in}$ and $\hat{a}_\mathrm{out}$, respectively,
	\begin{align}
	 \hat{a}_\mathrm{out}=\sqrt{\eta}\,\hat{a}_\mathrm{in}+\sqrt{1-\eta}\,\hat{c}_\mathrm{in},\label{Eq:IOR_operators}
	\end{align}
	where $\hat{c}_\mathrm{in}$ is the operator of the environment mode being in the vacuum state and $\eta\in[0,1]$ is the transmittance efficiency.
	This input-output relation can be easily transfered to the corresponding relations for the density operators of the input and attenuated modes, $\hat{\rho}_\mathrm{in}$ and $\hat{\rho}_\mathrm{att}(\eta)$, respectively\cite{mandel_book}.
	The density operator of the latter mode depends on the channel transmittance.
	
	A special feature of atmospheric channels consists in the fact that the transmittance $\eta$ is a random variable\cite{semenov09}.
	Hence, the quantum-state input-output relation for free-space channels requires the additional averaging over this random variable with respect to the PDT $\mathcal{P}(\eta)$,
	\begin{align}
	 \hat{\rho}_\mathrm{out}=\int\limits_{0}^{1}\D\eta\mathcal{P}(\eta)\hat{\rho}_\mathrm{att}(\eta).\label{Eq:IOR_DensityOperators}
	\end{align}
	Specifying the representation, we can get quantum-state input-output relations, which are convenient for different applications.
	For example, in the Glauber-Sudarshan representation\cite{glauber63c,sudarshan63}, this relation is given by
	\begin{align}
	 P_\mathrm{out}(\alpha)=\int\limits_{0}^{1}\D\eta\mathcal{P}(\eta)\frac{1}{\eta}P_\mathrm{in}\left(\frac{\alpha}{\sqrt{\eta}}\right),\qquad C_\mathrm{out}(\beta)=\int\limits_{0}^{1}\D\eta\mathcal{P}(\eta)C_\mathrm{in}(\sqrt{\eta}\beta),\label{Eq:IOR_P}
	\end{align}
	where $P_\mathrm{in/out}(\alpha)$ is the Glauber-Sudarshan $P$ function of the input/output mode and $C_\mathrm{in/out}(\beta)$ is the corresponding characteristic function.
	Note that the Glauber-Sudarshan $P$ function fully describes the corresponding quantum state.
	In the Wigner representation, the input-output relation reads as,
	\begin{align}
	  W_\mathrm{out}(\alpha)=\int\limits_{0}^{1}\D\eta\mathcal{P}(\eta)\frac{1}{\eta}\int_\mathbb{C}\D^2\gamma\, W_\mathrm{in}\left(\frac{\gamma}{\sqrt{\eta}}\right)\frac{2}{\pi \eta(1-\eta)}\exp\left[-\frac{2|\alpha-\gamma|^2}{1-\eta}\right].\label{Eq:IOR_W}
	\end{align}
	Here $W_\mathrm{in/out}(\alpha)$ is the Wigner function\cite{wigner32} of the input/output mode.
	Also, the input-output relation for the normally-ordered moments is of special interest
	\begin{align}
	 \langle\hat{a}^{\dag n}\hat{a}^m\rangle_\mathrm{out}=\langle\eta^{(m+n)/2}\rangle\langle\hat{a}^{\dag n}\hat{a}^m\rangle_\mathrm{in},\label{Eq:IOR_Mom}
	\end{align}
	where $\langle\eta^{(m+n)/2}\rangle=\int_{0}^{1}\D\eta \mathcal{P}(\eta) \eta^{(m+n)/2}$.
	
	The form of the PDT $\mathcal{P}(\eta)$ depends on beam parameters, propagation distance, aperture at the output site, and of course on the meteorological conditions.
	Under clear weather conditions, fluctuations of the transmittance are mostly caused by fluctuations of the part of the beam transmitted through the aperture.
	In this case, absorption and scattering lead to additional deterministic, i.e. non-fluctuating, losses.
	Different models for these fluctuations\cite{vasylyev12,vasylyev16,vasylyev18} have been proposed.
	It is also important to note that atmospheric precipitations may result in additional fluctuations of the beam form, which significantly effects the PDT\cite{vasylyev17}.
	Important features of the PDT take place in the case of the Cassegrain aperture, which is considered in the companion paper\cite{VasylyevSPIE}.

\section{PHOTOCOUNTING STATISTICS}
\label{Sec:PCS}

\begin{figure}[ht!!]
		\begin{center}
		\includegraphics[width=0.5\textwidth]{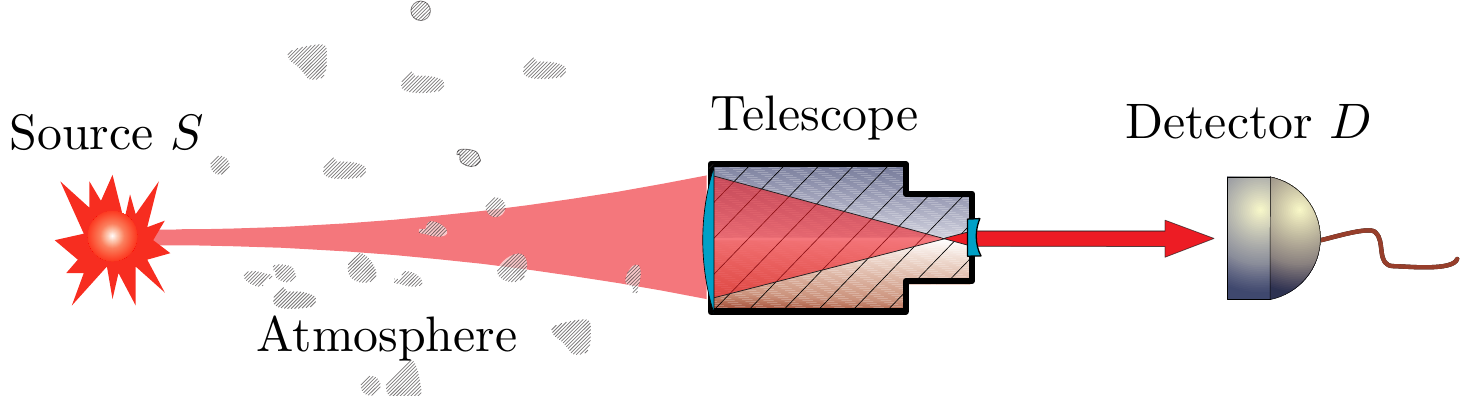}
		\end{center}
		\caption{\label{Fig:Photocounting} A sketch of the photocounting experiment for the light passed through the turbulent atmosphere.
		At the receiver site, the light is collected by a telescope and sent to a photodetector.
		The photocurrent produced by the latter is considered to be related to the number of detected photons $n$.}
		\end{figure}

	The photocounting experiment was the first experiment which has been theoretically considered for light transmitted through the atmosphere \cite{perina73, milonni04}, see Fig. \ref{Fig:Photocounting}.
	In recent experimental implementations\cite{capraro12} such a measurement is discussed in the context of its application to the decoy-state protocol\cite{Hwang2003,Lo2205}.
	In general, the experimental setup consist of a light source at the transmitter site and a telescope connected to a detector at the receiver site.
	Due to atmospheric turbulence and beam broadening, only a part of the beam passes through the aperture of the receiver telescope.
	A photodetector measures the photonnumbers $n$ of the impinging light.
	The corresponding photocounting statistics is described by the photocounting equation\cite{mandel_book,kelley64},
		\begin{align}
		p_n=\Tr\left(\hat{\rho}_\mathrm{out}\hat{\Pi}_n\right)=\int\limits_\mathbb{C}\D^2\alpha P(\alpha)\Pi_n(\alpha),\label{Eq:PhEq}
		\end{align}
	where
		\begin{align}
		\hat{\Pi}_n(\alpha)=:\frac{(\eta_c\hat{n}+\nu)^n}{n!}\exp\left(-\eta_c\hat{n}-\nu\right):\label{Eq:POVM_PNR}
		\end{align}
	is the positive operator-valued measure (POVM) for counting $n$ photons and
		\begin{align}
		\Pi_n(\alpha)=\frac{(\eta_c|\alpha|^2+\nu)^n}{n!}\exp\left(-\eta_c|\alpha|^2-\nu\right)
		\end{align}
	is its $Q$-symbol.
	Here $\eta_c$ is the detector efficiency, $\nu$ is the mean number of noise counts related to detector dark and stray light\cite{Semenov2008,Lee2005,Pratt1969,Karp1970}, and $\hat{n}=\hat{a}^\dag\hat{a}$ is the photon-number operator.

	In cases where the Glauber-Sudarshan $P$ function is not positive semidefinite (nonclassical)\cite{titulaer65}, the photocounting statistics may have a so-called sub-Poissonian character, which means that $\langle\Delta n^2\rangle<\langle n\rangle$.
	Here, $\langle\Delta n^2\rangle=\langle n^2\rangle-\langle n\rangle^2$ is the variance of the photocounting statistics.
	This property is characterized by the Mandel $Q$ parameter\cite{mandel_book,mandel79},
		\begin{align}
		Q=\frac{\langle\Delta n^2\rangle}{\langle n\rangle}-1.\label{Eq:Mandel}
		\end{align}
	For the sub-Poissonian light this parameter is negative, i.e. $Q<0$.

	Let us consider light being distributed through the atmospheric channel.
	At the transmitter site its Mandel parameter is $Q_\mathrm{in}$.
	The Mandel parameter at the receiver site is given by
		\begin{align}
		Q_\mathrm{out}=\frac{\langle\eta^2\rangle\langle n\rangle_\mathrm{in}}{\langle\eta\rangle\langle n\rangle_\mathrm{in}+\nu}Q_\mathrm{in}+\langle\Delta\eta^2\rangle\frac{\langle n\rangle_\mathrm{in}^2}{\langle\eta\rangle\langle n\rangle_\mathrm{in}+\nu}.\label{Eq:IOR_Mandel}
		\end{align}
	This input-output relation for the Mandel parameter includes two terms\cite{semenov09,bohmann17a}.
	The first one resembles the same for channels with deterministic (constant) losses.
	The second one appears due to the presence of fluctuating losses, i.e. atmospheric turbulence.
	From this relation, one can conclude that for intense light, such that
		\begin{align}
		\langle n\rangle_\mathrm{in}\geq-\frac{\langle\eta^2\rangle}{\langle\Delta\eta^2\rangle}Q_\mathrm{in}
		\end{align}
	the light at the receiver is not characterized by sub-Poissonian statistics.
	It is also important that this bound does not depend on the intensity of dark counts and stray light, $\nu$.
	Therefore, this disturbing factor can diminish the amount of negativity of the Mandel parameter but it does not result in its vanishing.

\section{BELL NONLOCALITY}
\label{Sec:Bell}

\begin{figure}[ht!!]
		\begin{center}
		\includegraphics[width=0.7\textwidth]{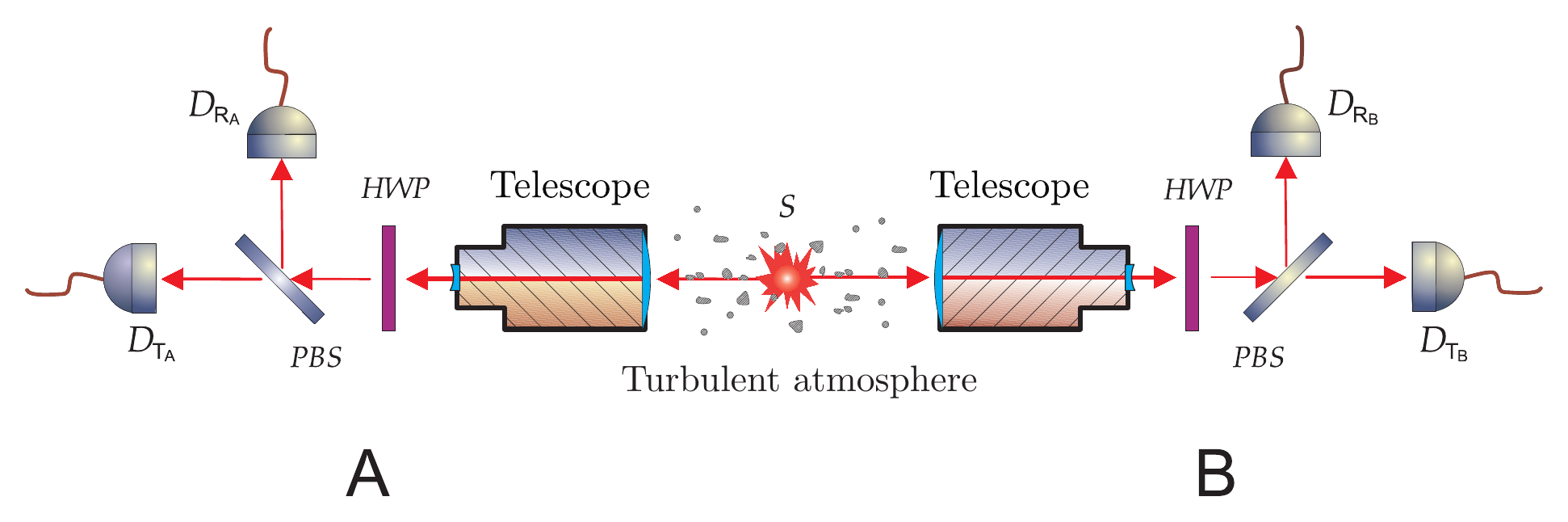}
		\end{center}
		\caption{\label{Fig:Bell-setup} An experimental setup for testing the Bell inequalities.
		The source irradiates light in two different directions. After passing through the atmospheric channel it is collected by telescopes and analyzed by polarization analyzers. See the text for more details.}
		\end{figure}

	Bell nonlocality\cite{Brunner2014} is the key resource for the Ekert protocol E91\cite{ekert91} as well as for device-independent quantum key distribution\cite{Acin2006,Acin2007}.
	A typical experimental setup is sketched in Fig. \ref{Fig:Bell-setup}.
	Light in a specially prepared polarization state is irradiated in two different directions---to the receiver sites A and B.
	After the collection by telescopes the light is analyzed by polarization analyzers.
	Each polarization analyzer consists of a half-wave plate (HWP), which rotates the polarization by the angle $\theta_\mathrm{A}$($\theta_\mathrm{B}$); a polarization beam splitter (PBS), which transmits the horizontally-polarized light and reflects the vertically-polarized light; and two detectors for the transmitted and reflected channel, $D_{T_{A(B)}}$ and $D_{R_{A(B)}}$, respectively.

	We consider the probability $P_{i_\mathrm{A},i_\mathrm{B}}(\theta_\mathrm{A},\theta_\mathrm{B})$ of simultaneous clicks of the detectors $i_A=\{T_\mathrm{A},R_\mathrm{A}\}$ and $i_B=\{T_\mathrm{B},R_\mathrm{B}\}$.
	In the most ideal situation, the state of light at the source is specially designed such that not more than one detector at each side is clicking.
	However, double click events at each site are still possible due to the presence of multiphoton pairs\cite{Ma2007}, detector dark counts, and stray light.
	For such events we assign a random bit\cite{Semenov2011,Beaudry2008,Moroder2010,Fung2011}.
	The measured probabilities are used to calculate the correlation coefficients for the given polarization angles
		\begin{align}
		E\left(\theta_\mathrm{A}, \theta_\mathrm{B}\right) =
		\frac{P_\mathrm{same}\left(\theta_\mathrm{A},
		\theta_\mathrm{B}\right)-P_\mathrm{different}\left(\theta_\mathrm{A},
		\theta_\mathrm{B}\right)}{P_\mathrm{same}\left(\theta_\mathrm{A},
		\theta_\mathrm{B}\right)+P_\mathrm{different}\left(\theta_\mathrm{A},
		\theta_\mathrm{B}\right)},
		\end{align}
	with
		\begin{align}
		&P_\mathrm{same}\!\left(\theta_\mathrm{A},\theta_\mathrm{B}\right){=}P_{T_\mathrm{A},
		T_\mathrm{B}}\!\left(\theta_\mathrm{A},\theta_\mathrm{B}\right)+P_{R_\mathrm{A},
		R_\mathrm{B}}\!\left(\theta_\mathrm{A},\theta_\mathrm{B}\right),\label{Eq:Psame}\\
		&P_\mathrm{different}\!\left(\theta_\mathrm{A},\theta_\mathrm{B}\right){=}P_{T_\mathrm{A},
		R_\mathrm{B}}\!\left(\theta_\mathrm{A},\theta_\mathrm{B}\right)+P_{R_\mathrm{A},
		T_\mathrm{B}}\!\left(\theta_\mathrm{A}, \theta_\mathrm{B}\right).
		\label{Eq:Pdifferent}
		\end{align}
	These coefficients with different settings of the polarization angles are substituted in an expression for the Bell parameter in the Clauser-Horn-Shimony-Holt (CHSH) form\cite{CHSH}
		\begin{align}
		\mathcal{B}=\left|E\left(\theta_\mathrm{A}^{(1)},
		\theta_\mathrm{B}^{(1)}\right)-E\left(\theta_\mathrm{A}^{(1)},
		\theta_\mathrm{B}^{(2)}\right)\right|
		+\left|E\left(\theta_\mathrm{A}^{(2)},
		\theta_\mathrm{B}^{(2)}\right)+E\left(\theta_\mathrm{A}^{(2)},
		\theta_\mathrm{B}^{(1)}\right)\right|\label{Eq:BellParameter}.
		\end{align}
	In local realistic theories this parameter cannot exceed the value of 2, i.e.
		\begin{align}
		\mathcal{B}\leq 2.\label{Eq:BellIneq}
		\end{align}
	In quantum theory this inequality can be violated.

	A theoretical analysis of the Bell inequality test in atmospheric channels\cite{semenov10,gumberidze16} results in the following expressions for the probabilities $P_i\left(\theta_\mathrm{A},\theta_\mathrm{B}\right)$, $i=\{\textrm{same},\textrm{different}\}$,
		\begin{align}
		P_\mathrm{i}\left(\theta_\mathrm{A},
		\theta_\mathrm{B}\right)&=\frac{1}{2}+\frac{e^{-4\nu}}{2}
		\left(1-\tanh^2\xi\right)^4\label{Eq:ProbabilitySpecialSame2}\\
		&{\times}\Bigg[e^{2\nu}\bigg(2\left\langle
		\frac{1}{C_\mathrm{0}+C_\mathrm{1A}+C_\mathrm{1B
		}+C_\mathrm{i}}\right\rangle
		{-}\left\langle\frac{C_\mathrm{0}}{{(C_\mathrm{0}+C_\mathrm{1A})}^2}\right\rangle{-}
		\left\langle\frac{C_\mathrm{0}}{{(C_\mathrm{0}+C_\mathrm{1B})}^2}\right\rangle\nonumber\\
		&-2\left\langle\frac{1}{C_\mathrm{0}+C_\mathrm{1A}+C_\mathrm{1B}+C_\mathrm{
		j}}\right\rangle\bigg){+}\left\langle\frac{1}{C_\mathrm{0}}\right\rangle\Bigg]
		.\nonumber
		\end{align}
	Here $i{\neq}j$,
		\begin{align}
		C_\mathrm{0}=\left\{\eta_c^2\eta_\mathrm{A}\eta_\mathrm{B}\tanh^2\xi-
		\left[1+\left(\eta_c \eta_\mathrm{A}-1\right)\tanh^2\xi\right]\left[1+\left(\eta_c
		\eta_\mathrm{B}-1\right)\tanh^2\xi\right]\right\}^2,\label{Eq:C02}
		\end{align}
		\begin{align}
		C_\mathrm{1A(B)}&=\eta_c \eta_\mathrm{B(A)}\left(1-\eta_c
		\eta_\mathrm{A(B)}\right)\left(1-\tanh^2\xi\right)\tanh^2\xi\label{Eq:C1A}
		\\
		&\times\left\{\eta_c^2\eta_\mathrm{A(B)}\eta_\mathrm{B(A)}\tanh^2\xi-
		\left[1+\left(\eta_c
		\eta_\mathrm{A(B)}-1\right)\tanh^2\xi\right]\left[1+\left(\eta_c
		\eta_\mathrm{B(A)}-1\right)\tanh^2\xi\right]\right\},\nonumber
		\end{align}
		\begin{align}
		C_\mathrm{same}=\eta_c^2
		\eta_\mathrm{A} \eta_\mathrm{B}\tanh^2\xi\left(1-\tanh^2\xi\right)
		^2\label{Eq:Csame2}
		\left[\left(1-\eta_c \eta_\mathrm{A}\right)\left(1-\eta_c \eta_\mathrm{B}\right)
		\tanh^2\xi-\sin^2
		\left(\theta_\mathrm{A}-\theta_\mathrm{B}\right)\right],
		\end{align}
		\begin{align}
		C_\mathrm{different}=\eta_c^2
		\eta_\mathrm{A} \eta_\mathrm{B}\tanh^2\xi\left(1-\tanh^2\xi\right)
		^2\label{Eq:Cdifferent2}
		\left[\left(1-\eta_c \eta_\mathrm{A}\right)\left(1-\eta_c \eta_\mathrm{B}\right)
		\tanh^2\xi-\cos^2
		\left(\theta_\mathrm{A}-\theta_\mathrm{B}\right)\right],
		\end{align}
	and $\xi$ is the squeezing parameter characterizing the presence of multiphoton pairs in the state at the receiver site.	
	The averaging with the atmospheric-channel transmittances $\eta_\mathrm{A}$ and $\eta_\mathrm{B}$ is given by
		\begin{align}
		\left\langle\ldots\right\rangle=\int\limits_{0}^{1}\D
		\eta_\mathrm{A}\D\eta_\mathrm{B}\ldots
		\mathcal{P}\left(\eta_\mathrm{A},\eta_\mathrm{B}\right),\label{Eq:mean}
		\end{align}
	where the the two-dimensional generalization of the PDT $\mathcal{P}\left(\eta_\mathrm{A},\eta_\mathrm{B}\right)$ is used.

\begin{figure}[ht!!]
		\begin{center}
		\includegraphics[width=0.4\textwidth]{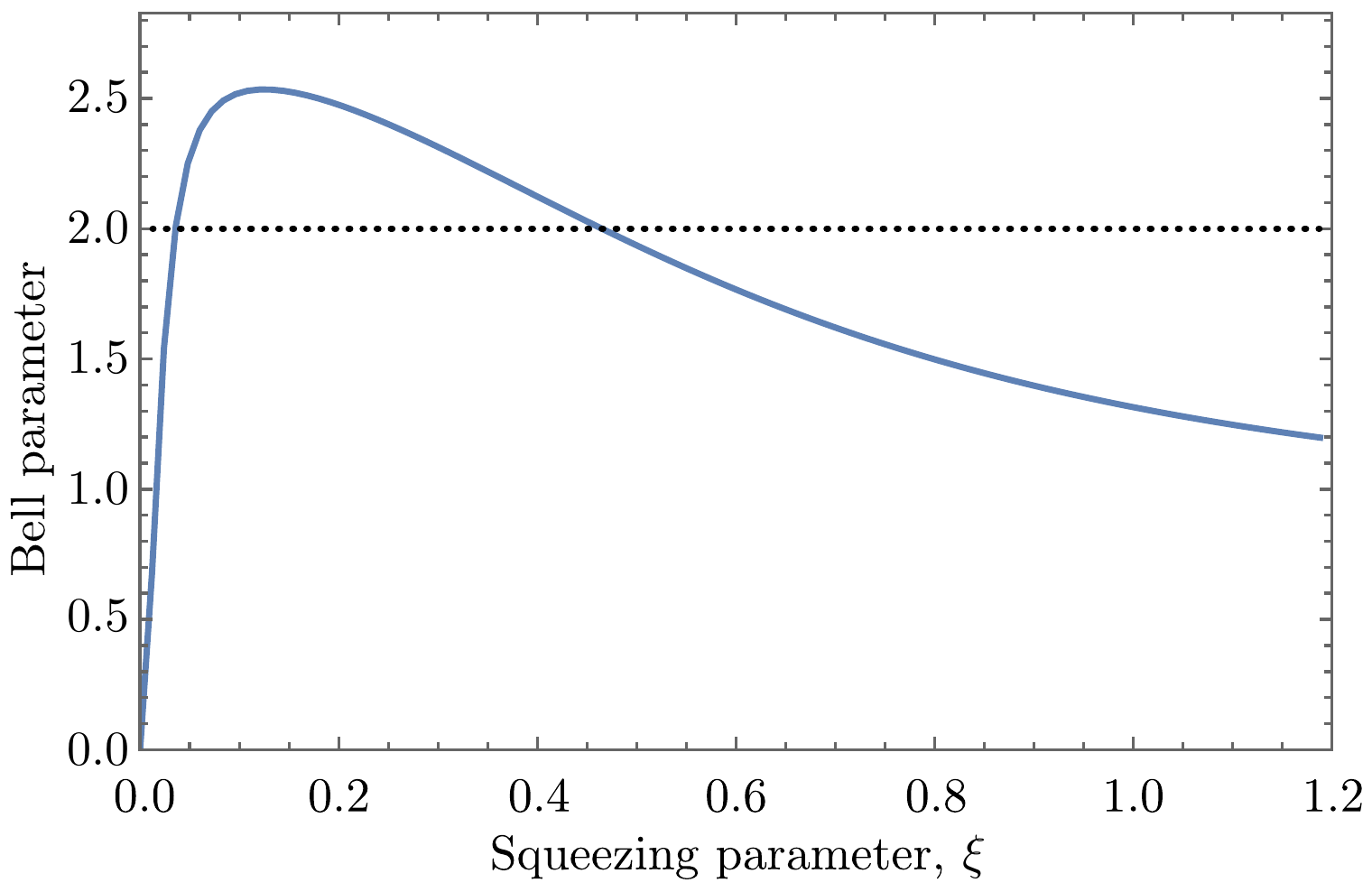}
		\end{center}
		\caption{\label{Fig:Counter-Bell} Bell parameter vs the squeezing parameter for the scenario of counterpropagation for two uncorrelated atmospheric channels\cite{VasylyevSPIE} of 4 km length each.
		The mean number of stray-light and dark counts is $\nu=5\times 10^{-4}$, the atmospheric refractive-index structure constant is $C_n^2=10^{-14}\textrm{m}^{-2/3}$, the detection efficiency is $\eta_c=0.6$, and the deterministic losses due to atmospheric absorption is $\eta_a=0.12$.
		The detection module  is assumed to be equipped with a 6-inch f/12 refractor telescope.}
		\end{figure}

	In the most typical scenario of counterpropagation, the channel transmittances are uncorrelated such that $\mathcal{P}(\eta_\mathrm{A},\eta_\mathrm{B})=\mathcal{P}(\eta_\mathrm{A})\mathcal{P}(\eta_\mathrm{B})$.
	As it has been demonstrated\cite{gumberidze16}, the Bell parameter in this case has practically the same values for atmospheric channels and for deterministic-loss channels with the efficiencies $\langle\eta_\mathrm{A(B)}\rangle$.
	In Fig.~\ref{Fig:Counter-Bell} the dependence of the Bell parameter on the squeezing parameter is given.
	A pronounced minimum near the zero-point of the squeezing parameter is explained by vanishing contributions of the nonclassical source in comparison to detector dark counts and stray light.
		
\begin{figure}[ht!!]
		\begin{center}
		\includegraphics[width=0.4\textwidth]{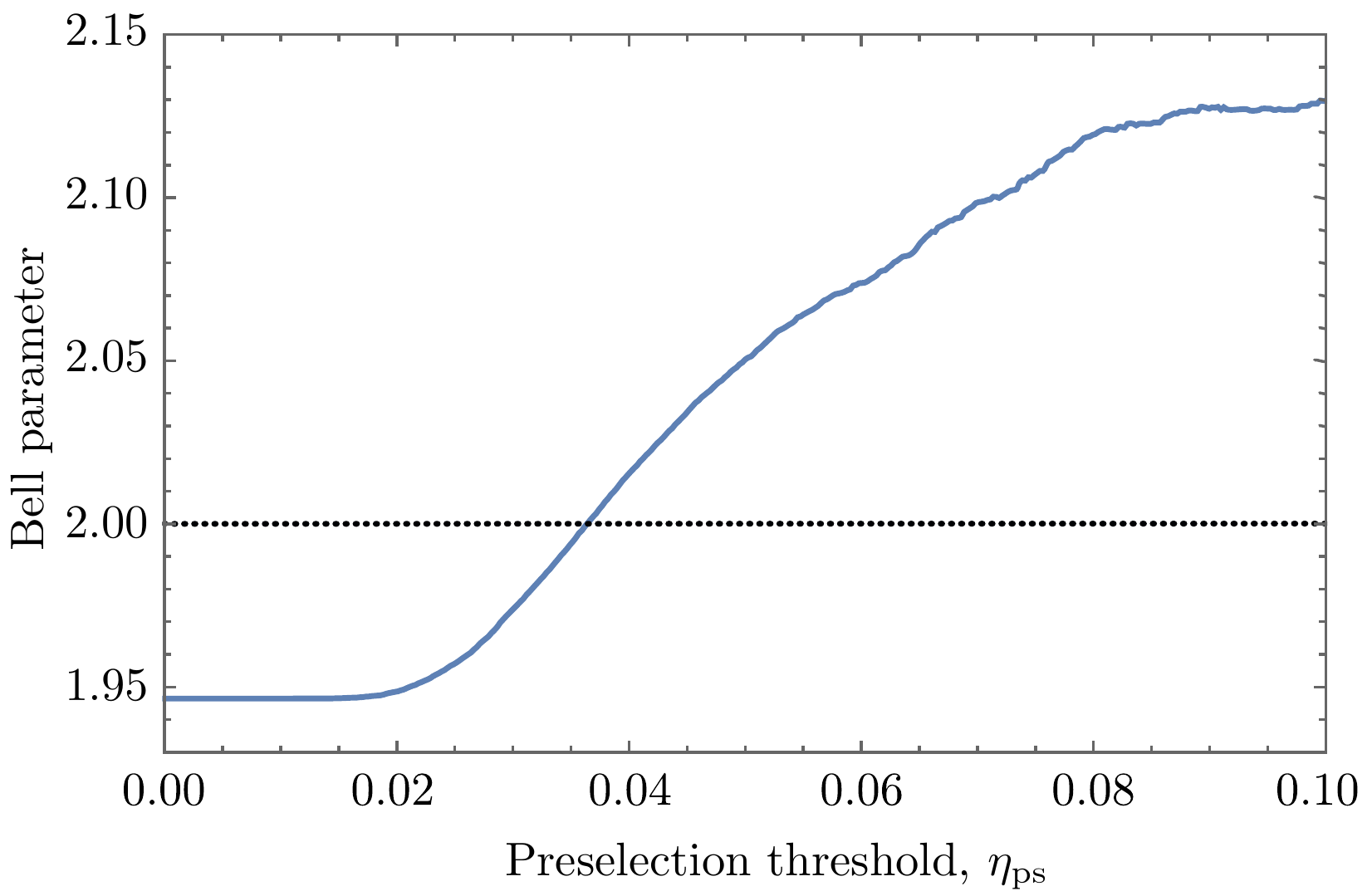}
		\end{center}
		\caption{\label{Fig:Preselection} Bell parameter vs the preselection threshold $\eta_\mathrm{ps}$ in the scenario of counterpropagation for the same channels as in Fig.~\ref{Fig:Counter-Bell} but for $\nu=2\times 10^{-3}$.}
		\end{figure}

	In the scenario of counterpropagation, a scheme which we refer to as preselection appears to be useful.
	This uses the idea of testing the atmospheric channels with strong light pulses and was proposed by Capraro et al.\cite{capraro12}
	The properties of the atmosphere change very slowly\cite{Tatarskii} in comparison to times between pulses used in quantum-optical experiments.
	For that reason, the measured value of the transmittance for the strong reference pulse can be considered to be the same as for the subsequent nonclassical pulses during a certain time interval.
	We preselect only the events, whose transmittance appears to be more than a certain preselection threshold $\eta_\mathrm{ps}$.
	In Fig. \ref{Fig:Preselection}, the dependence of the Bell parameter on the preselection threshold for 4km atmospheric channel is shown.
	From this plot, we can conclude that even in the case were the direct verification of Bell nonlocality is impossible due to strong dark counts and stray light, the preselection procedure enables us to obtain the value of the Bell parameter larger than the classical bound $\mathcal{B}=2$.

	Another possibility related to the verification of Bell nonlocality is the scenario of copropagation, which has been experimentally implemented for a 144km channel at the Canary Islands by Fedrizzi et al.\cite{fedrizzi09}
	For such a scenario, both light modes are sent in the same direction but with a small time delay.
	In this case the channels can be considered as correlated such that the two-dimensional PDT is given by $\mathcal{P}(\eta_\mathrm{A}, \eta_\mathrm{B})=\mathcal{P}(\eta_\mathrm{A})\delta(\eta_\mathrm{A}-\eta_\mathrm{B})$, where $\delta$ is the delta distribution.
	In this case, the probability that two photons will be detected is $\langle\eta_\mathrm{A}^2\rangle$.
	The probability that two photons are detected in the deterministic loss channels with the efficiency $\eta_0=\langle\eta_\mathrm{A}\rangle$ is $\langle\eta_\mathrm{A}\rangle^2$.
	Applying the Cauchy-Schwartz inequality, $\langle\eta_\mathrm{A}^2\rangle\geq\langle\eta_\mathrm{A}\rangle^2$, we conclude that the probability of detecting both photons in the atmospheric channels can be more than the corresponding probability for the deterministic channels.
	
\begin{figure}[ht!!]
		\begin{center}
		\includegraphics[width=0.4\textwidth]{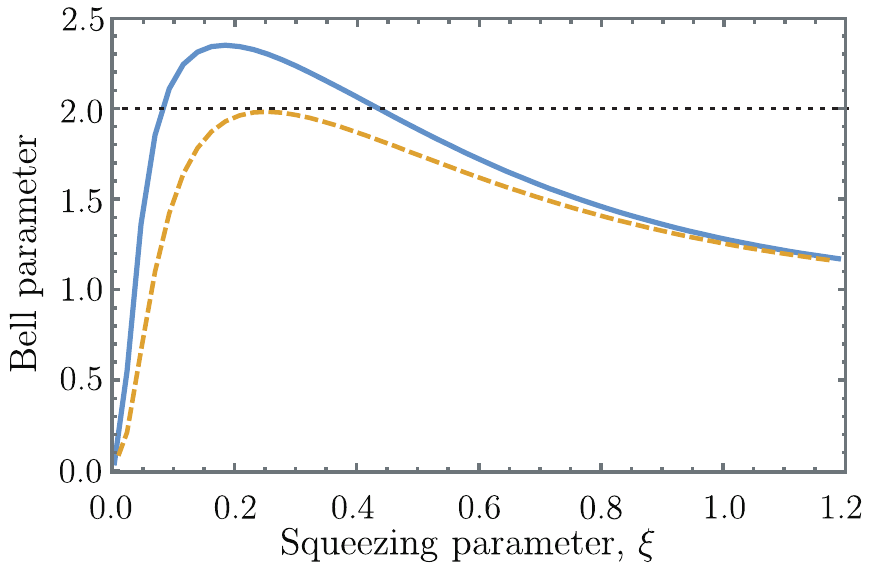}
		\end{center}
		\caption{\label{Fig:Co-Bell} Bell parameter vs the squeezing parameter in the scenario of copropagation for a 144km atmospheric channel in the Canary Islands for $\nu=1.7\times10^{-5}$.
		The PDT is considered in the form of truncated log-normal distribution with the parameters measured by Capraro et. al.\cite{capraro12}
		Solid and dashed lines correspond to the atmospheric and deterministic-loss channels, respectively.
		Adopted from the paper by Gumberidze et. al.\cite{gumberidze16}}
		\end{figure}

	This behavior is demonstrated in Fig. \ref{Fig:Co-Bell} for the channel between the Canary Islands.
	For the corresponding intensity of the dark counts and stray light there is no violation of Bell inequalities for deterministic-loss channel of the same mean transmittance.
	In the fluctuating-loss channel the violation appears to be clearly verified.
	However, we have not found a significant difference between channels with deterministic and fluctuating losses in the case of low and average mean losses\cite{gumberidze16}.

\section{QUADRATURE SQUEEZING}
\label{Sec:Sqeez}

	In this section, we consider quantum effects based on measurements of continuous-valued observables.
	A typical example of measurements for the corresponding continuous-variable (CV) protocols is the balanced homodyne detection, which is used for measuring the field quadrature $ \hat{x}(\varphi)=\frac{1}{\sqrt{2}}\left(\hat{a} e^{-i\varphi}+\hat{a}^\dag e^{i\varphi}\right)$\cite{mandel_book,vogel_book, VogelReview}.
	For the special cases $\hat{x}(0)=\hat{q}$ and $\hat{x}(\pi/2)=\hat{p}$, the quadratures play the role similar to the operators of position and momentum in quantum mechanics.

	Balanced homodyne detection is based on combining the signal field with a strong laser signal---the local oscillator.
	The phase between the two fields should be synchronized.
	A typical solution for this synchronization consists in generating the signal and the local oscillator with a common source.
	For atmospheric channels, such a solution meets several problems since the local oscillator should also be sent through the atmosphere, which leads to chaotic changes of its amplitude and phases.

\begin{figure}[ht!!]
			\begin{center}\includegraphics[width=0.8\linewidth]{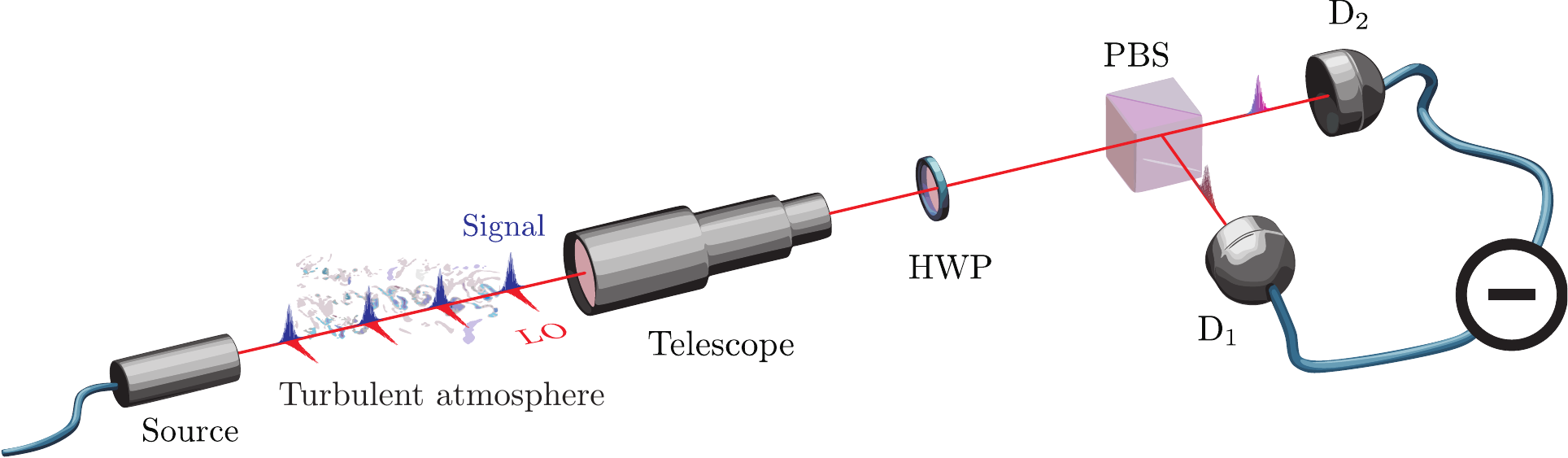}\end{center}
			\caption{\label{Fig:HDScheme}
				Scheme of homodyne detection for quantum light passing through the turbulent atmosphere as proposed by Elser et al.\cite{elser09}, and Heim et al.\cite{heim10}.
			See the text for more explanations.}
		\end{figure}

	An elegant solution of this problem has been proposed by Elser et al.\cite{elser09} and Heim et al.\cite{heim10}.
	The scheme of the corresponding experiment is sketched in Fig. \ref{Fig:HDScheme}.
	The idea is that the signal and the local oscillator are sent in the same spatial but orthogonally polarized modes.
	The depolarization effect of the atmosphere is negligible small\cite{Tatarskii}.
	For this reason, the relative phase $\varphi$ appears to be almost perfectly stabilized.
	The signal and the local oscillator are collected by the same telescope and combined using a half-wave plate (HWP) and a polarization beam splitter (PBS).
	Detectors at the outputs of the PBS produce the photocurrents $I_1$ and $I_2$.
	Then, the value of the field quadrature is $x(\varphi)\sim(I_1-I_2)/\sqrt{I_1+I_2}$.

	A theoretical analysis of this measurement scheme has been proposed by Semenov et al.\cite{semenov12}
	Data processing in the presence of detector dark counts and stray light necessarily includes so-called postprocessing noise.
	In fact, this is equal to the statement that the imperfect data processing result in the appearance of an effective density operator, which differs from the density operator of the state at the receiver site.
	The corresponding effective $P$ function, $P_\mathrm{out}\!\left(\alpha\right)$, is given by
		\begin{align}
		P_\mathrm{out}\!\left(\alpha\right)=\int\limits_{0}^{1}\D
		\eta\,\mathcal{P}\!\left(\eta\right) \frac{1}{\eta}
		\exp\left[\frac{\nu}{4 r^2
			\eta^{2}}\,\Delta_\alpha\right]
		P\!\left(\frac{\alpha}{\sqrt{\eta}}\right),
		\label{Eq:P_Monit_Noisy2}
		\end{align}
	where $r$ is the amplitude of the local oscillator, $\nu$ is the mean intensity of detector dark counts and stray light, and $\Delta_\alpha=\frac{\partial^2}{\partial (\re\alpha)^2}+\frac{\partial^2}{\partial (\im\alpha)^2}$.
	In many practical situations, however, the channel transmittance is bounded by its minimal value $\eta_\mathrm{min}$.
	For such a scenario the local-oscillator amplitude can be chosen such that the condition  $r^2\gg\nu/\eta_\mathrm{min}^2$ is satisfied.
	In this case, Eq. (\ref{Eq:P_Monit_Noisy2}) is reduced to the standard input-output relation for the atmospheric channels (\ref{Eq:IOR_P}).

	The variance of the quadrature, $\langle \Delta \hat{q}^2\rangle$, can be expressed in terms of its normal-ordered form, $\langle :\Delta \hat{q}^2:\rangle$, as $\langle \Delta \hat{q}^2\rangle=\langle: \Delta \hat{q}^2:\rangle+{1}/{2}$.
	For the states $\langle :\Delta \hat{q}^2:\rangle<0$ the quadrature noise measured with the balanced homodyne detection is less than the corresponding noise of the vacuum state.
	Such states are referred to as quadrature squeezed\cite{Stoler1970,Stoler1971,Yuen1976,Wu1986,Wu1987,Vahlbruch}.
	Quadrature squeezing is a resource for some CV quantum-key distribution protocols\cite{Hillery2000,Cerf2001,Gottesman2001,Madsen2012}.

\begin{figure}[ht!!]
			\begin{center}\includegraphics[width=0.55\linewidth]{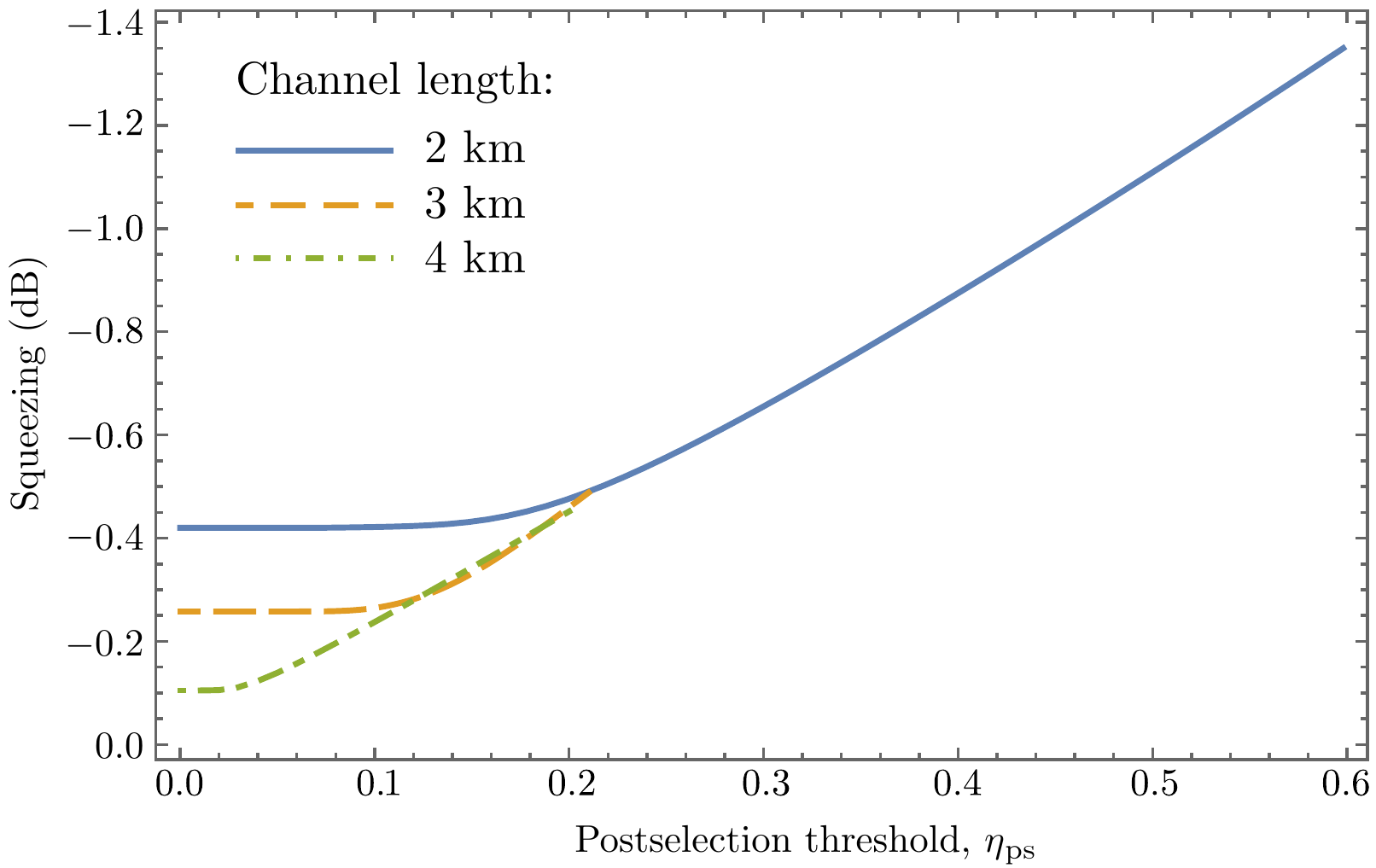}\end{center}
			\caption{\label{Fig:SqueezingPostselection}
				The value of squeezing in dB vs. the postselection threshold, $\eta_\mathrm{ps}$, for different channels discussed in the companion paper by Vasylyev et al.\cite{VasylyevSPIE}.
				Initially the state is prepared with -2.4~dB squeezing.}
		\end{figure}
	
	Applying the input-output relations in the form of Eq. (\ref{Eq:IOR_Mom}) one gets the input-output relation for the squeezing certifier, $\langle :\Delta \hat{q}^2:\rangle$,
		\begin{align}
		\langle: \Delta \hat{q}^2:\rangle_\textrm{out}=\langle T^2\rangle\langle: \Delta \hat{q}^2:\rangle_\textrm{in}+\langle\Delta T^2\rangle\langle\hat{q}\rangle_\mathrm{in}^2,\label{Eq:IOR_Squeezing}
		\end{align}
	where $T=\sqrt{\eta}$ is the channel transmission coefficient.
	Similar to Eq. (\ref{Eq:IOR_Mandel}), this relation includes two terms.
	The first one resembles the standard attenuation, while the second one describes the effect of atmospheric turbulence.
	From this equation one can conclude that the effect of turbulence destroys quadrature squeezing in the case of nonzero coherent displacement, $\langle\hat{q}\rangle_\mathrm{in}$, in the direction of the squeezed quadrature.
	The absence of such a displacement is the best scenario for transmitting the quadrature squeezing through the atmosphere.

	Since the signal intensity is much smaller than the intensity of the local oscillator, the channel transmittance is proportional to the sum of the photocurrents on the detectors, $\eta\sim I_1^2+I_2^2$.
	By controlling this transmittance, one can implement a postselction procedure.
	This means that after the measurement of the pair quadrature-transmittance, $\{x(\varphi),\eta\}$, we select only the events with transmittances larger than a postselection threshold $\eta_\mathrm{ps}$.
	Such an experiment has been implemented by Peuntinger et al.\cite{peuntinger14} for the 1.6km atmospheric channel in the city of Erlangen.
	The experimental results are in good agreement with the elliptic-beam model for the atmospheric channel\cite{vasylyev16}.
	In Fig. \ref{Fig:SqueezingPostselection}, the dependence of the squeezing on the postselection threshold for channels of different length is shown.
	The data are obtained with the PDT model based on the law of total probabilities\cite{vasylyev18}, for details see the companion paper\cite{VasylyevSPIE}.
	One can see that up to some extend the increasing of the propagation length enables one to preserve quadrature squeezing if the postselection procedure is applied.

\section{GAUSSIAN ENTANGLEMENT}
\label{Sec:Entanglement}

	Entanglement\cite{horodecki09} is one of the most fundamental quantum properties and a key resource for many quantum communication protocols.
	The most general definition of entanglement has been given by Werner\cite{werner89}.
	According to the definition a two-mode state $\hat{\rho}$ is separable if and only if there exist a decomposition into single-mode states, $\hat{\rho}_\mathrm{A}^{(k)}$ and $\hat{\rho}_\mathrm{B}^{(k)}$, such that $\hat{\rho}=\sum_k p_k\hat{\rho}_\mathrm{A}^{(k)}\otimes\hat{\rho}_\mathrm{B}^{(k)}$, with $p_k\geq0$ and $\sum_kp_k=1$.
	If the state cannot be presented in such a form, it is entangled.
	According to the Peres-Horodecki separability criterion\cite{peres96,horodecki96} the state $\hat{\rho}$ is entangled if its partial transposition is not a positive semidefinite operator.

	Two-mode Gaussian states can be completely characterized by first and second moments of the annihilation and creation operators for each mode, $\hat{a}$, $\hat{b}$ and $\hat{a}^\dag$, $\hat{b}^\dag$.
	In that case the Peres-Horodecki criterion is a necessary and sufficient condition of the entanglement.
	It can be reformulated in the form of the Simon criterion\cite{simon00}.
	According to the Simon criterion in the form given by Shchukin and Vogel\cite{shchukin05c} the state is entangled if and only if
		\begin{align}
		\mathcal{W}=\det V^\mathrm{PT}<0,\label{Eq:SimonDuanDeterm}
		\end{align}
		where $ V^\mathrm{PT}$ is the partial transposition of the matrix
		\begin{align}\label{Eq:SimonMatrix}
			\begin{aligned}
				V{=}
				\begin{pmatrix}
					\langle \Delta\hat a^\dagger\Delta \hat a\rangle & \langle \Delta\hat a^{\dagger2}\rangle & \langle \Delta\hat a^\dagger\Delta \hat b\rangle & \langle \Delta\hat a^\dagger\Delta \hat b^\dagger\rangle \\
					\langle \Delta\hat a^2\rangle & \langle \Delta\hat a\Delta\hat a^\dagger\rangle & \langle \Delta\hat a\Delta \hat b\rangle & \langle \Delta\hat a\Delta \hat b^\dagger\rangle \\
					\langle \Delta\hat a\Delta\hat b^\dagger\rangle & \langle \Delta\hat a^\dagger\Delta\hat b^\dagger\rangle & \langle \Delta\hat b^\dagger\Delta \hat b\rangle & \langle \Delta \hat b^{\dagger2}\rangle \\
					\langle \Delta\hat a\Delta\hat b\rangle & \langle \Delta\hat a^\dagger\Delta\hat b\rangle & \langle \Delta\hat b^2\rangle & \langle \Delta \hat b\Delta \hat b^\dagger\rangle
				\end{pmatrix}
			\end{aligned}.
		\end{align}
	Similarly, the less general Duan-Giedke-Cirac-Zoller (DGCZ) criterion\cite{Duan2000} states only a sufficient condition for Gaussian entanglement.
	In this case,the matrix $V$  in Eq. (\ref{Eq:SimonDuanDeterm}) is replaced by
		\begin{align}\label{Eq:DuanMatrix}
				V=\begin{pmatrix}
					\langle\Delta \hat{a}^\dagger \Delta \hat{a} \rangle
					& \langle\Delta \hat{a}^\dagger \Delta \hat{b} \rangle
					\\ \langle\Delta \hat{a} \Delta \hat{b}^\dagger \rangle
					& \langle\Delta \hat{b}^\dagger \Delta \hat{b} \rangle
				\end{pmatrix},
		\end{align}
	which is a $2\times2$ sub-matrix of the matrix (\ref{Eq:SimonMatrix}).

	After passing through the atmospheric channel, the quantum state of light is not Gaussian in general.
	However, we can consider only its Gaussian part.
	Gaussian entanglement is important and can be considered as a resource for, e.g., the Braunstein-Kimble CV-teleportation protocol\cite{Braunstein1998}.

	The general input-output relation for the Simon certifier (\ref{Eq:SimonDuanDeterm}) has been considered in the paper by Bohmann et al.\cite{bohmann16}.
	Here, we discuss a less general but also important case of the DGCZ criterion.
	An important motivation for such a restriction is the fact that the DGCZ-entangled states always preserve entanglement after passing through the deterministic-loss channels\cite{barbosa11}.
	This is explained by the fact that the corresponding lossy DGCZ certifier is a rescaling of the input certifier,
	\begin{align}
	 \mathcal{W}_{\eta_a,\eta_b}=\eta_a\eta_b \mathcal{W}_\mathrm{in},\label{Eq:DuanAtt}
	\end{align}
	where $\eta_a$ and $\eta_b$ are transmittances of the channels for the modes $a$ and $b$, respectively.
	For atmospheric channels, the input-output relation for the DGCZ certifier is given by
		\begin{align}
		\mathcal{W}_\textrm{out}=\mathcal{W}_{\langle T_{a}^2\rangle,\langle T_{b}^2\rangle}+N+\boldsymbol{\nu}^\dag S\boldsymbol{\nu}+\boldsymbol{\mu}^\dag F\boldsymbol{\mu}.\label{Eq:DuanIOR}
		\end{align}
	Similar to Eq. (\ref{Eq:IOR_Mandel}) for the Mandel parameter and Eq. (\ref{Eq:IOR_Squeezing}) for the squeezing certifier, the first term of this equation corresponds to the deterministic-loss channels with the transmittances $\langle T_{a}^2\rangle$ and $\langle T_{b}^2\rangle$.
	Here and in the following $T_{a,b}=\sqrt{\eta_{a,b}}$ are the corresponding transmission coefficients.
	The next terms appear as the effect of atmospheric turbulence.
	Two of these terms depend on the coherent displacements of the input state via $\boldsymbol{\nu}=\big(\big\langle\hat{a}\big\rangle,\big\langle\hat{b}^\dag\big\rangle\big)^\mathrm{T}$ and $\boldsymbol{\mu}=\big(\big\langle\hat{a}\big\rangle\big\langle\hat{b}\big\rangle,\big\langle\hat{a}^\dag\big\rangle\big\langle\hat{b}^\dag\big\rangle\big)^\mathrm{T}$.
	This is an important difference of the fluctuating-loss channels from the deterministic-loss channels.
	The coefficient $N$ and the $2\times2$ matrices $S$ and $F$ are given by
		\begin{align}
		N=\left(\big\langle T_a^2\big\rangle\big\langle
		T_b^2\big\rangle-\big\langle T_a T_b\big\rangle^2\right)
		\big\langle\Delta\hat{a}\Delta\hat{b}\big\rangle
		\big\langle\Delta\hat{a}^\dag\Delta\hat{b}^\dag\big\rangle,\label{Eq:N}
		\end{align}
		\begin{align}
		 S=
			\begin{pmatrix}
			\big\langle T_b^2\big\rangle\big\langle\Delta\hat{b}^\dag\Delta\hat{b}\big\rangle
			&-\big\langle T_a T_b\big\rangle\big\langle\Delta\hat{a}\Delta\hat{b}\big\rangle\\
			-\big\langle T_a T_b\big\rangle\big\langle\Delta\hat{a}^\dag\Delta\hat{b}^\dag\big\rangle
			&\big\langle T_a^2\big\rangle\big\langle\Delta\hat{a}^\dag\Delta\hat{a}\big\rangle
			\end{pmatrix}\circ
			\begin{pmatrix}
			\big\langle \Delta T_a^2\big\rangle
			&\big\langle \Delta T_a \Delta T_b\big\rangle\\
			\big\langle \Delta T_a \Delta T_b\big\rangle&
			\big\langle \Delta T_b^2\big\rangle
			\end{pmatrix},\label{Eq:S}
		\end{align}
		\begin{align}
		 F=\frac{1}{2}\left(\big\langle \Delta T_a^2 \big\rangle\big\langle \Delta T_b^2
		\big\rangle-\big\langle \Delta T_a\Delta T_b\big\rangle^2\right)
		\begin{pmatrix}
		1&0\\0&1
		\end{pmatrix},\label{Eq:F}
		\end{align}
	where $\circ$ means the entrywise Hadamard-Schur product of matrices.
	The input-output relation for the Simon certifier\cite{bohmann16} has the same form but with different $N$, $S$, and $F$.
	
	In the standard scenario of counterpropagation, the channels are in general uncorrelated.
	This means that in Eqs. (\ref{Eq:N}), (\ref{Eq:S}), and (\ref{Eq:F}) one should put $\big\langle \Delta T_a \Delta T_b\big\rangle=0$ and $\big\langle T_a T_b\big\rangle=\big\langle T_a\big\rangle \big\langle T_b\big\rangle$.
	It is easy to check that in this case $N\geq0$, and the matrices $S$ and $F$ are positive semidefinite.
	Hence, the contribution of atmospheric turbulence increases the value of the DGCZ certifier.
	The latter is also increased by increasing the coherent displacement of the input field.
	Therefore, the most optimal scenario for sending Gaussian entanglement is the case of $\boldsymbol{\nu}=0$, i.e., for zero displacement.
	
	Let us consider the case of counterpropagation for the two-mode squeezed vacuum state (TMSV),
		\begin{align}
		\ket{\rm TMSV}=\frac{1}{\cosh\xi}\sum_{n=0}^{\infty}\tanh^n \xi \ket{n,n}.\label{Eq:TMSV}
		\end{align}
	Here $\xi$ is the squeezing parameter.
	For this state, a counterintuitive result\cite{bohmann16} is the fact that large values of squeezing can destroy the Gaussian entanglement at the receiver site.
	
	Another scenario is the case of correlated channels.
	This can be implemented via a copropagation scheme similar to those realized by Fedrizzi et al.\cite{fedrizzi09} for testing Bell-inequalities.
	Also correlation of transmittances can be established for counterpropagating modes.
	In this case, one can apply the adaptive scheme proposed in the paper by Bohmann et al.\cite{bohmann16}.
	For this purpose both parties should measure their transmittances and share the information via classical communication.
	The party whose transmittance appears to be larger attenuates its channel to the level of the lower one.
	In such a way we obtain correlated losses by cost of additional losses.
	
	Advantages of the correlated channels can be seen by analyzing the input-output relation (\ref{Eq:DuanIOR}).
	The correlation conditions $T_a=T_b\equiv T$ implies that $N=0$, $F=0$, and $S$ is not a positive semidefinite matrix.
	Then the input-output relation for the DGCZ certifier reads
		\begin{align}\label{Eq:IOR_DuanCorr}
			\mathcal{W}_\textrm{out}
			=\langle T^2\rangle^2\mathcal{W}_\textrm{in}
			+\langle\Delta T^2\rangle \langle T^2\rangle
			\begin{pmatrix}\langle\hat a\rangle\\\langle\hat b^\dagger\rangle\end{pmatrix}^\dagger
			\begin{pmatrix}
				\langle\Delta\hat{b}^\dag\Delta\hat{b}\rangle
				&-\langle\Delta\hat{a}\Delta\hat{b}\rangle
				\\ -\langle\Delta\hat{a}^\dag\Delta\hat{b}^\dag\rangle
				& \langle\Delta\hat{a}^\dag\Delta\hat{a}\rangle
			\end{pmatrix}
			\begin{pmatrix}\langle\hat a\rangle\\\langle\hat b^\dagger\rangle\end{pmatrix}.
		\end{align}
	In absence of coherent displacements the second term in this input-output relation vanishes and the DGCZ certifier at the receiver site has exactly the same form as for the case of deterministic-loss channels.
	This means that the DGCZ-entangled states without coherent displacement always remain entangled after passing through correlated channels.
	
		\begin{figure}[t]
			\begin{center}\includegraphics[width=0.8\linewidth]{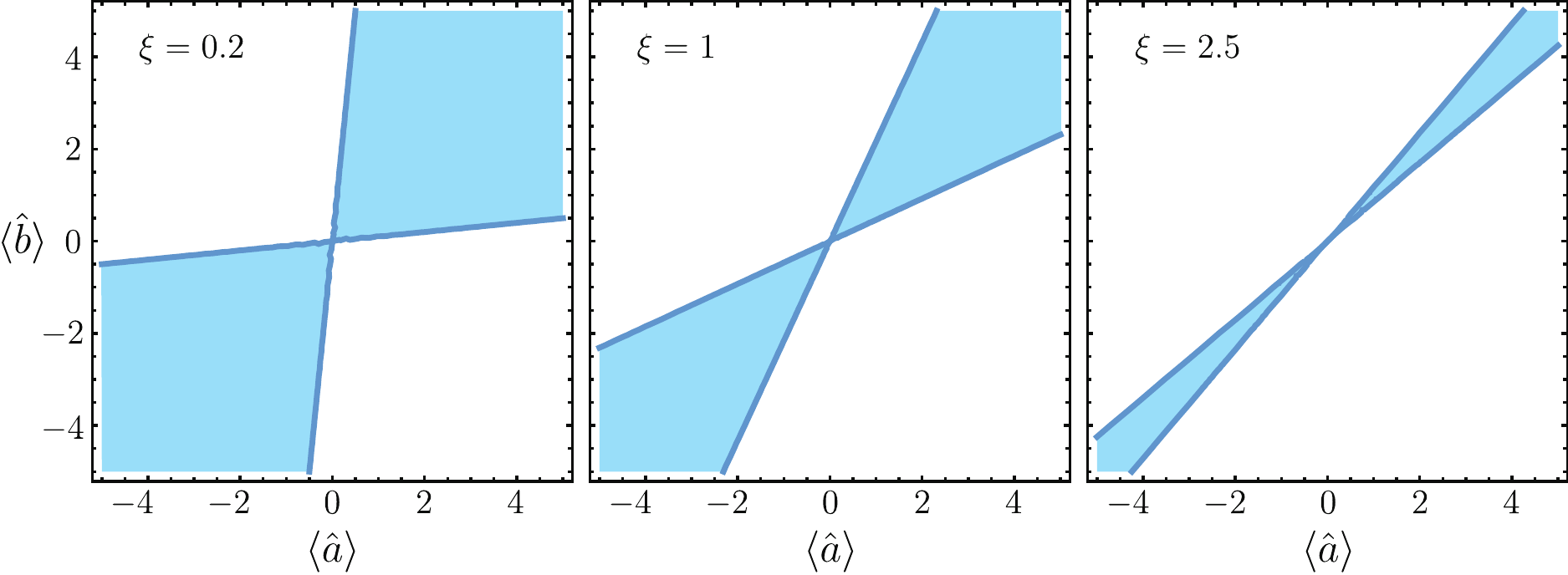}\end{center}
			\caption{\label{Fig:Duan} The shaded area indicates the domain in the plane of real coherent displacements, $\{\langle\hat a\rangle,\langle\hat b\rangle\}$, where the entanglement is preserved independently on the state of the atmosphere for different squeezing parameters.}
		\end{figure}
	
	Since the matrix $S$ for such channels is not positive definite, entanglement for the correlated channels can be preserved even for large values of coherent amplitudes $\langle\hat a\rangle$ and $\langle\hat b\rangle$.
	Moreover, for the DGCZ-entangled states there exists a domain of their values for which entanglement is always preserved, independent of the state of the atmosphere.
	This domain is determined by non-positive values of the second term in Eq. (\ref{Eq:IOR_DuanCorr}),
		\begin{align}
			\begin{pmatrix}\langle\hat a\rangle\\\langle\hat b^\dagger\rangle\end{pmatrix}^\dagger
			\begin{pmatrix}
				\langle\Delta\hat{b}^\dag\Delta\hat{b}\rangle
				&-\langle\Delta\hat{a}\Delta\hat{b}\rangle
				\\ -\langle\Delta\hat{a}^\dag\Delta\hat{b}^\dag\rangle
				& \langle\Delta\hat{a}^\dag\Delta\hat{a}\rangle
			\end{pmatrix}
			\begin{pmatrix}\langle\hat a\rangle\\\langle\hat b^\dagger\rangle\end{pmatrix}\leq0.
		\end{align}
	For example, for the TMSV state (\ref{Eq:TMSV}) this condition is given by
		\begin{align}
		 \sinh^2(\xi)\left(|\langle\hat a\rangle|^2+|\langle\hat b\rangle|^2\right)-\sinh(2\xi)\re\left(\langle\hat a\rangle\langle\hat b\rangle\right)\leq0.
		\end{align}
	In Fig. \ref{Fig:Duan}, this domain is shown for the case of real-valued coherent displacements, $\{\langle\hat a\rangle,\langle\hat b\rangle\}$.
	It is interesting to note that with increasing squeezing parameter $\xi$ this domain is decreasing.
	Hence, in this case strong values of squeezing are not desirable.

\section{SUMMARY AND CONCLUSIONS}\label{sec:conclusions}
\label{Sec:Concl}

	To conclude, we note that atmospheric quantum channels can be used to transfer different nonclassical properties of light.
	We presented input-output relations, which connect the quantum states at the transmitter site to the one on the receiver site.
	This mathematical tool appears to be useful for analyzing a variety of quantum communication schemes.
	At the same time, it is worth to note that in many situations applications of additional techniques, such as postselction or correlation strategies, are useful for preserving nonclassical properties of light.
	In Tab. \ref{Tab:concl} we give a brief summary of the applicability of such methods.

		\begin{table}[h!!!]
		\caption{\label{Tab:concl} Applicability of different techniques for preserving nonclassical effects in atmospheric quantum channels are shown.}
		\begin{center}
		\begin{tabular}{|c|c|c|c|c|}
		\hline
		{ }&Postselection&Preselection& Adaptive correlations&Copropagation\\
		\hline
		Sub-Poissonian statistics&na&{\color{Green}\ding{52}}&na&na\\
		\hline
		Bell nonlocality&na&{\color{Green}\ding{52}}&{\color{red}\ding{56}}&{\color{Green}\ding{52}}\\
		\hline
		Quadrature squeezing&{\color{Green}\ding{52}}&&na&na\\
		\hline
		Gaussian entanglement &{\color{Green}\ding{52}}&&{\color{Green}\ding{52}}&{\color{Green}\ding{52}}\\
		\hline
		\end{tabular}
		\end{center}
		\end{table}

	CV quantum communication protocols are based on the procedure of balanced homodyne detection.
	This technique, among other, enables a real-time control of the channel transmittance.
	By postselecting the events with transmittance exceeding the threshold, we obtain a possibility of preserving quadrature squeezing and Gaussian entanglement.
	This procedure can also be used for preserving higher-order nonclassicality and entanglement.
	
	In discrete-variable protocols, real time control of the channel transmittance is possible by sending strong classical pulses before the pulses of nonclassical light.
	In this case, we can preselect the measurements of such events, which correspond to large values of channel transmittance.
	Particularly, we showed that this is useful for preserving Bell nonlocality under conditions of strong background radiation.
	Furthermore, it became clear that the preselection procedure can be successfully applied for preserving nonclassical statistics of photocounts.
	
	The main advantage of adaptive channel correlations consists of the fact that with this technique we do not need to apply any post- or pre-selection procedures.
	This adaptive method preserves or establishes channel correlations by cost of additional losses.
	The technique appears to be useful for preserving Gaussian entanglement.
	However, we did not find examples when it can significantly improve the verification of Bell nonlocality.
	
	Some communication protocols use two nonclassically-correlated modes of light which are sent in the same direction.
	This can be a very promising technique for atmospheric links since in this case both channels are almost perfectly correlated.
	Particularly, we have shown that in such a scenario, Bell nonlocality can be preserved even better than for comparable deterministic-loss channels.
	This technique also always preserves entanglement of DGCZ-entangled states.
	
	Free space channels have very important differences from deterministic optical-fiber based loss channels.
	Their features can be used for developing novel quantum communication schemes.
	We are convinced that the physical ideas, discussed in this paper, will be useful in the corresponding communication engineering applications.

	\acknowledgments
	The authors acknowledge enlightening discussions with Stefan Gerke, Jan Sperling, and Maria Gumberidze. 
	The work was supported by the Deutsche Forschungsgemeinschaft through projects VO 501/21-2 and VO 501/22-2.


\end{document}